\documentclass[pra,aps,twocolumn,groupedaddress,nofootinbib]{revtex4-2} %superscriptaddress
\usepackage{graphicx}
\usepackage[nointegrals]{wasysym} %
\usepackage[export]{adjustbox}
\usepackage{amsmath,amsfonts,amssymb,latexsym}
\usepackage{hhline}
\usepackage{bm}
\usepackage{verbatim}
\usepackage{enumitem}
\usepackage{mathrsfs}
\usepackage{slashed}
\usepackage{empheq}
\usepackage{physics}

\usepackage{graphicx}
\usepackage{dcolumn}
\usepackage{bm}
\usepackage{multirow}

\usepackage[normalem]{ulem}

\usepackage{amsmath}
\usepackage{amsthm}
\usepackage{amstext}
\usepackage{amssymb}
\usepackage{mathrsfs}
\usepackage{amsfonts}
\usepackage{amsbsy} 

\usepackage{braket}

\usepackage[all,matrix,cmtip]{xy}
\usepackage{mathtools}

\usepackage{xcolor}
\definecolor{red}{rgb}{1,0,0}
\definecolor{blue}{rgb}{0,0,1}
\definecolor{dblue}{rgb}{0,0,0.4}
\definecolor{green}{rgb}{0,1,0}
\definecolor{black}{rgb}{0,0,0}
\definecolor{white}{rgb}{1,1,1}
\definecolor{pastelblue}{RGB}{20,93,160}

\definecolor{brn}{rgb}{.8,.4,.0}
\definecolor{redo}{rgb}{1,.5,.0}
\definecolor{ddgrn}{rgb}{0,0.4,0}
\definecolor{dgrn}{rgb}{0,0.55,0}
\definecolor{dbl}{rgb}{0,0,0.5}

\usepackage[colorlinks,citecolor=pastelblue,linkcolor=pastelblue,urlcolor=pastelblue]{hyperref}

\newcommand{\bpm}{\begin{pmatrix}}
	\newcommand{\epm}{\end{pmatrix}}
\newcommand{\bmm}{\begin{matrix}}
	\newcommand{\emm}{\end{matrix}}
\newcommand{\bvm}{\begin{vmatrix}}
	\newcommand{\evm}{\end{vmatrix}}

\usepackage{euscript}

\makeatletter
\newsavebox{\@brx}
\newcommand{\llangle}[1][]{\savebox{\@brx}{\(\m@th{#1\langle}\)}%
	\mathopen{\copy\@brx\kern-0.5\wd\@brx\usebox{\@brx}}}
\newcommand{\rrangle}[1][]{\savebox{\@brx}{\(\m@th{#1\rangle}\)}%
	\mathclose{\copy\@brx\kern-0.5\wd\@brx\usebox{\@brx}}}
\makeatother

\allowdisplaybreaks

\newcommand{\bs}{\boldsymbol}
\usepackage{booktabs}
\usepackage{tabularx}
\usepackage{array}

\begin{document}

%\preprint{APS/123-QED}

\title{Superconductivity and non-Fermi liquid metals in a charge-1/3 anyon fluid}  

\author{Zhengyan Darius Shi}%
\email{zhengyanshi@stanford.edu}
\affiliation{
Leinweber Institute for Theoretical Physics, Stanford University, Stanford, California 94305, USA}

\author{T. Senthil}
\email{senthil@mit.edu}
 %\homepage{}
\affiliation{
Department of Physics, Massachusetts Institute of Technology,
Cambridge, Massachusetts 02139, USA
}%

\date{\today}

\begin{abstract}

We revisit the charge-1/3 anyon fluid obtained by doping the $\nu = 2/3$ Jain fractional Chern insulator (FCI).  In the standard composite fermion description, the doped anyons fractionalize into three translation-related flavors of secondary composite fermions, whose gauge-mediated interactions drive a robust inter-flavor pairing instability. In our previous work, we analyzed a flavor-asymmetric paired state and obtained a charge-ordered Fermi liquid. Inspired by a recent paper, we consider an alternative flavor-symmetric paired state and show that it is an SC* state: a charge-$2e$ condensate that coexists with residual $\mathbb{Z}_2$ topological order. The weak and strong pairing regimes share the same intrinsic topological order but differ in chiral central charge, giving $c_- = 7/2$ and $c_- = 2$. We further show how other proposed effective field theories fit within the same composite fermion description, and argue that across the doping driven FCI-to-superconductor transition, localized anyons evolve into Bogoliubov quasiparticles rather than vortices. At low doping, we identify an approximate SU(3)-symmetric regime in which the system instead realizes a non-Fermi liquid $\mathbb{Z}_3$ Orthogonal Metal with three charge-1/3 fermion pockets and no sharp electron quasiparticle. Finally, we comment on the energetics of various possible ground states and discuss implications for experiments in moire materials. 

% We revisit the charge-1/3 anyon fluid obtained by doping the $\nu = 2/3$ Jain fractional Chern insulator (FCI). Motivated by the observation of zero field Jain-sequence states and composite Fermi liquids in moire materials, we analyze the doped system within the standard composite fermion description. At nonzero doping, the composite fermions fractionalize into three translation-related flavors of secondary composite fermions, whose gauge-mediated interactions drive a robust inter-flavor pairing instability. For the flavor-symmetric paired state, we find that the resulting phase is not an ordinary topological superconductor but an SC* state: a charge-$2e$ condensate that coexists with residual $\mathbb{Z}_2$ topological order. The weak and strong pairing regimes share the same intrinsic topological order but differ in chiral central charge, giving $c_- = 7/2$ and $c_- = 2$ . We show how other proposed effective field theories and the resulting superconductors can be incorporated into the standard composite fermion description. We argue that  across the doping driven FCI-to-superconductor transition, localized anyons evolve into Bogoliubov quasiparticles rather than vortices. Finally, we identify an approximate SU(3)-symmetric regime in which the doped system instead realizes a non-fermi liquid $\mathbb{Z}_3$ orthogonal metal with three charge-1/3 fermion pockets and no sharp electron quasiparticle. Together, these results broaden the landscape of possible superconducting and metallic phases in doped FCIs.
\end{abstract}

\maketitle

\section{Introduction}
%The discovery of Fractional Quantum Anomalous Hall (FQAH) insulators~\cite{Cai2023_FQAHTMD,Park2023_FQAH_TMD,Xu2023_FQAHTMD,Zeng2023_FQAHTMD,Lu2023_FQAHPenta,Lu2025_EQAH} (or more broadly Fractional Chern Insulators (FCI)~\cite{Sun2011_FCI,Sheng2011_FCI,Regnault2011_FCI,Tang2011_FQAH,Wang2011_FQAH,Neupert2010_FQAH,Bergholtz2013_FCIreview,Parameswaran2013_FCIreview,Spanton2017_FCI,Xie2021_FCI,Aronson2024_FCI,Aronson2025_FCI_penta,Butler2026_FCI_rhombo}) has re-invigorated\cite{Shi2024_doping,Divic2024_HofHubb, Kim2024_anyonSC,Zhang2025_SU(3)1dope,Pichler2025_anyonSC,Shi2025_dopeMR,Nosov2025_plateau,Shi2025_anyon_delocalization,Han2025_anyonexciton,Nakajima2025_thermo_anyon,Kuhlenkamp2025_HofHubb,Shi2025_nonAbelian_TSC,lotrivc2026phases,Fan2026_weakpairing_SC,Wang2026_U3anyonSC} the study of possible ground states of a inite-density anyon gas~\cite{Laughlin1988_anyonSC,Lee1989_anyonSC,Fetter1989_anyonSC_RPA,Chen1989_anyonSC,Wen1990_anyonSC,Tang2013_anyonSC}. 
Unlike in  the usual high field realizations of fractional quantum Hall phenomena, the charged anyonic excitations of the recently discovered Fractional Quantum Anomalous Hall~\cite{Cai2023_FQAHTMD,Park2023_FQAH_TMD,Xu2023_FQAHTMD,Zeng2023_FQAHTMD,Lu2023_FQAHPenta,Lu2025_EQAH} (or more broadly Fractional Chern Insulators (FCI)~\cite{Sun2011_FCI,Sheng2011_FCI,Regnault2011_FCI,Tang2011_FQAH,Wang2011_FQAH,Neupert2010_FQAH,Bergholtz2013_FCIreview,Parameswaran2013_FCIreview,Spanton2017_FCI,Xie2021_FCI,Aronson2024_FCI,Aronson2025_FCI_penta,Butler2026_FCI_rhombo}) states are {\em mobile anyons} with a band dispersion~\cite{Shi2024_doping,Schleith2025_anyondisp,Goncalves2025_anyondisp,Yan2025_anyondisp,Iyer2026_anyondisp,wang2026measuring}.  Doping these insulators thus generates a fluid of fractionally charged anyons~\cite{Laughlin1988_anyonSC,Lee1989_anyonSC,Fetter1989_anyonSC_RPA,Chen1989_anyonSC,Wen1990_anyonSC,Tang2013_anyonSC}, potentially leading to exotic itinerant phases~\cite{Shi2024_doping,Divic2024_HofHubb, Kim2024_anyonSC,Zhang2025_SU(3)1dope,Pichler2025_anyonSC,Shi2025_dopeMR,Nosov2025_plateau,Shi2025_anyon_delocalization,Han2025_anyonexciton,Nakajima2025_thermo_anyon,Kuhlenkamp2025_HofHubb,Chen2025_SCdopeHH,Shi2025_nonAbelian_TSC,lotrivc2026phases,Fan2026_weakpairing_SC,Wang2026_U3anyonSC}. 

Motivated by this prospect, in our earlier work~\cite{Shi2024_doping}, we described a number of possible ground states obtained by doping FQAH insulators within an effective field theory framework. For the prominent $2/3$ state seen in current experiments, we showed that doping the charge-$2/3$ anyon naturally leads to a charge-$2e$ BCS superconductor (with no coexisting topological order) and chiral central charge $c_- = -2$ ({\it i.e.} $4$ counter-propagating Majorana edge modes at the sample boundary); we will label this state as SC$_{-2}$ where the subscript denotes the chiral central charge.  

%In Ref. \cite{Shi2024_doping}, we also initiated the study of the possible ground states obtained by doping  the charge-$1/3$ anyon  using the standard composite fermion (CF) formalism~\cite{Jain1989_CFframework,Lopez1991_CSGL,Kol1993_Jainlattice,Halperin1993_HLRtheory}. %, implemented through a parton construction. At small dopings away from $2/3$, we showed that the CFs live in 3 valleys related by lattice translation and, in each valley, see an emergent  magnetic field corresponding to Landau level filling $1/2$. A natural possibility then is that each CF valley forms a Fermi surface of ``secondary" CFs \footnote{The terminology means these are composite fermions formed out of the basic primary composite fermions through a further flux attachment}, thereby leading to a secondary composite Fermi liquid (CFL) metal with 3 fermi pockets. We showed that this metal is unstable to pairing of the secondary CFs. A natural proposal for the resulting ground state--obtained by interpocket $p+ip$ pairing of just 2 pockets--is an ordinary Fermi liquid metal coexisting with a doping induced period-$3$ charge density wave. The Fermi volume of this metal is set by the dopant density. A more exotic possibility--obtained by $p+ip$ intra-pocket pairing of secondary CFs--is a Pair Density Wave superconductor coexisting with a non-Abelian topological order.

In Ref.~\cite{Shi2024_doping}, we also initiated the study of possible ground states obtained by doping in charge-$1/3$ anyons using the standard composite fermion (CF) formalism~\cite{Jain1989_CFframework,Lopez1991_CSGL,Kol1993_Jainlattice,Halperin1993_HLRtheory}.   
At low doping, the CFs live in 3 valleys related by lattice translation and, in each valley, see an emergent  
magnetic field corresponding to Landau level filling $1/2$. A natural possibility is that each CF further fractionalizes to form Fermi surfaces of secondary composite fermions (sCF) \footnote{The terminology means these are composite fermions formed out of the basic primary composite fermions through a further flux attachment}, thereby leading to a secondary composite Fermi liquid (sCFL) metal with 3 Fermi pockets. We showed that this metal is unstable to sCF pairing, and described some of the possible ground states.  
\begin{figure*}
    \centering
    \includegraphics[width=0.66\linewidth]{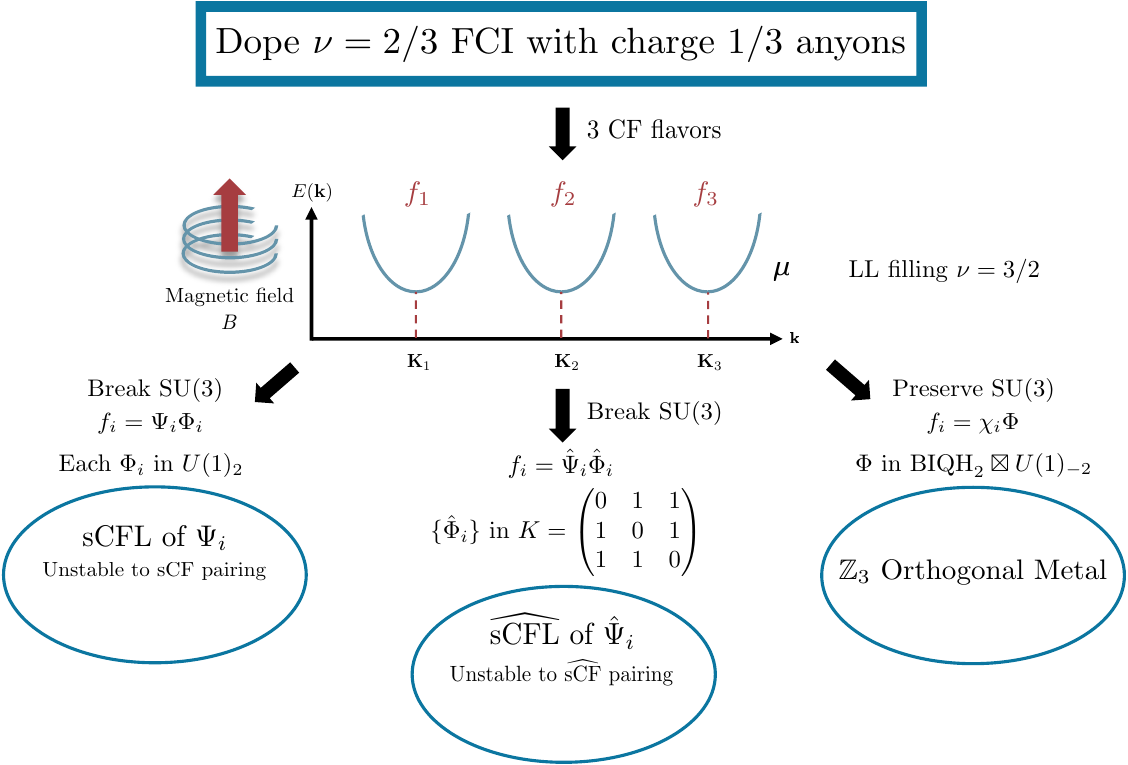}
    \caption{Summary of different families of itinerant phases that can arise from doping the $\nu = 2/3$ FCI with charge $1/3$ anyons under the standard CF framework. $U(1)_{\pm 2}$ is the bosonic Laughlin state at $\nu = \pm 1/2$, $\mathrm{BIQH}_2$ is the boson integer quantum Hall state at $\nu = 2$, and $K = (\ldots )$ is an Abelian topological order described by the corresponding $K$-matrix. Each family corresponds to a different fractionalization pattern for the doped CFs $f_i$. The sCFL and $\mathbb{Z}_3$ Orthogonal metal are analyzed in this work, while the $\widehat{\mathrm{sCFL}}$ is equivalent to the effective field theory of Ref.~\cite{Fan2026_weakpairing_SC} and the ``holon metal" of Ref.~\cite{Zhang2025_SU(3)1dope}.}
    \label{fig:Diff_sCF}
\end{figure*}

In subsequent work~\cite{Shi2025_dopeMR,Shi2025_nonAbelian_TSC}, we used alternative effective field theories to study ground states of $1/3$ anyons doped into the $2/3$ state to obtain translation-invariant charge-$2e$ topological superconductors SC$_{3/2}$ and SC$_{5/2}$. %In general all these effective field theories can be derived through parton constructions. Different such constructions lead to different effective field theories in terms of the corresponding partons, which for the present paper can be taken to be fermionic.  
Very recently, a different effective field theory~\cite{Fan2026_weakpairing_SC} (see also Ref.~\cite{Zhang2025_SU(3)1dope}) was developed for the charge-$1/3$ anyon fluid, and gave a route to a charge-$2e$ topological superconductor SC$_{-1/2}$. This theory is based on a fermionic  description of charge $1/3$ anyons that is different from the standard CFs discussed in our previous work. In general, all of these effective field theories can be derived through parton constructions. Which of these distinct possibilities obtains in a particular system is an energetic question that depends on details of the microscopic model. 
%In common with our work, Ref. \cite{Fan2026_weakpairing_SC} obtains 3 Fermi pockets of their fermionic partons (related again by lattice translation) which are unstable to pairing. A symmetric interpocket $p + ip$ ``weak" pairing of all 3 pockets leads to the $c_- = -1/2$ superconductor while the ``strong" pairing phase leads to the $c_- = -2$ superconductor (same as the one obtained by doping the charge $2/3$ anyons in Ref.~\cite{Shi2024_doping}). 

Given the observation of multiple Jain states and the Composite Fermi Liquid in tMoTe$_2$ (and in rhombohedral graphene), we expect that the standard CF framework provides a good starting point for capturing the energetics of the current experimental platforms. Using this framework, we obtain several new results in this paper which we summarize below. 

(1) To deal with the effective Landau level filling of the CFs in the doped state, we work within the sCF framework developed in Ref.~\cite{Shi2024_doping}. Inspired by Ref.~\cite{Fan2026_weakpairing_SC}, we describe the symmetric inter-flavor $p+ip$ paired state of these sCFs. At weak pairing, we find a charge-$2e$ SC$^*_{7/2}$ phase ({\it i.e}, a superconductor coexisting with topological order~\cite{Senthil2000_SC*}) coexisting with $\mathbb{Z}_2$ ``toric code" order.   At strong pairing, we obtain an SC$^*_{2}$ with the same topological order. 

(2) We show how the standard CF formalism can also lead to the ostensibly different effective theory of Refs.~\cite{Zhang2025_SU(3)1dope,Fan2026_weakpairing_SC}, if we fractionalize the standard CFs into an alternative set of secondary composite fermions, which we refer to as $\widehat{\mathrm{sCF}}$.

(3) At low doping density, the effective interaction between the standard CFs may have an approximate emergent $SU(3)$ symmetry. When this symmetry is present, neither the sCF framework of Ref.~\cite{Shi2024_doping} nor the $\widehat{\mathrm{sCF}}$ framework of Ref.~\cite{Fan2026_weakpairing_SC} is appropriate. We derive another effective theory that respects the SU(3) symmetry, and show that it leads to a fractionalized metal with 3 Fermi pockets of charge-$1/3$ fermions. This metal is a non-Fermi liquid with no sharp electron quasiparticle, and is similar to the Orthogonal Metals~\cite{nandkishore2012orthogonal} discussed previously as the simplest non-Fermi liquids. The sCF, $\widehat{\mathrm{sCF}}$ and SU(3) symmetric constructions are summarized in Fig.~\ref{fig:Diff_sCF}.

(4) We discuss the phenomenology associated with the superconductors obtained through the mechanism of {sCF/$\widehat{\mathrm{sCF}}$ pairing and contrast it with the phenomenology of doped charge 2/3 anyon fluids. Finally we comment on existing experiments and numerical results. 

A synthesis of the different possibilities discussed in the literature for the doped state near the 2/3 FQAH state is in Table \ref{tab:scnflnear2/3}. 

\begin{table*}
    \centering
    \begin{ruledtabular}
    \begin{tabular}{ccc}
        Doped state & Minimal dopant charge & Theoretical description \\
        $\mathrm{SC}_{-2}$ & 2/3  & Quantum Hall states of fermionic partons in a U(1) gauge theory~\cite{Shi2024_doping}  \\
        $\mathrm{SC}_{-1/2}$  & 1/3  & Symmetric inter-flavor pairing of fermionic partons~\cite{Fan2026_weakpairing_SC,Wang2026_U3anyonSC} \\
        $\mathrm{SC}_{3/2}$ & 1/3 & Quantum Hall states of bosonic partons in a U(2) gauge theory~\cite{Shi2025_nonAbelian_TSC} \\
        $\mathrm{SC}_{5/2}$ & 1/3  & Quantum Hall states of fermionic partons in a U(2) gauge theory~\cite{Shi2025_nonAbelian_TSC}\\
        $\mathrm{SC}^*_{7/2}$ & 1/3  & Symmetric inter-flavor pairing of fermionic partons (this work) \\
        PDW SC* & 1/3  & Intra-flavor pairing of fermionic partons~\cite{Shi2024_doping} \\
        CDW metal  & 1/3  & Pairing between two (out of three) flavors of fermionic partons~\cite{Shi2024_doping,Zhang2025_SU(3)1dope,Fan2026_weakpairing_SC} \\
        Z3OM non-Fermi liquid  &  1/3  & SU(3)-symmetric composite fermions (this work) \\
    \end{tabular}
    \end{ruledtabular}
    \caption{Some of the itinerant phases that can arise from doping the $\nu = 2/3$ Jain FCI. The subscript on SC refers to the chiral central charge. The SC$^*$ states have superconductivity coexisting with intrinsic topological order. We have only included charge-$2e$ superconductors. Charge-$4e$ superconductors are also in principle possible~\cite{Shi2025_nonAbelian_TSC,Shi2026_4eTQC}. The Z3OM is a non-Fermi liquid metal. All of these states can be obtained through distinct parton constructions.}
    \label{tab:scnflnear2/3}
\end{table*}

\section{Doping the Laughlin state with composite fermions}

To discuss the doped $2/3$ state, it is convenient to first do a particle-hole transformation and consider doping the $1/3$  Laughlin state. In the remainder of this paper, we will assume that the microscopic system is well-described by the standard CF mean field theory near $\nu = 1/3$. Working on the lattice, we obtain CF $f$ through a parton construction $c = f \Phi$ and put $\Phi$ in a gapped $U(1)_2$ bosonic Laughlin state. At $\nu = 1/3$, $f$ sees $1/3$ flux per unit cell and forms a Chern insulator with $C = 1$~\cite{Shi2024_doping}. At low doping, translation symmetry enforces three CF flavors $f_i$ in the reduced Brillouin zone, leading to an effective theory for the doped system~\footnote{Note that on an infinite plane, we can integrate out $\alpha$ to generate an improperly quantized Chern-Simons term $\frac{1}{8\pi} (A-a) d (A-a)$, which recovers the familiar form used in early literature~\cite{Lopez1991_CSGL,Halperin1993_HLRtheory}.}
\begin{equation*}
    L_{\rm doped} = \sum_{i=1}^3 L[f_i, a] + \mathrm{CS}[a,g] - \frac{2}{4\pi} \alpha d \alpha + \frac{1}{2\pi} \alpha d (A - a) \,,
\end{equation*}
where $\mathrm{CS}[a,g]$ is a short hand for $\frac{1}{4\pi} a d a + 2\mathrm{CS}_g$, which includes the integer quantum Hall response to the $\mathrm{spin}_{\mathbb{C}}$ connection $a$ as well as the background metric $g$.~\footnote{For reasons explained in this context in Ref.~\cite{Shi2024_doping}, in writing continuum field theories it will be convenient to include a coupling to a metric and to keep track of distinctions between $U(1)$ gauge fields and spin$_{\mathbb{C}}$ connections.} More precisely, microscopic lattice translations embed in the IR as a $\mathbb{Z}_3 \times \mathbb{Z}_3$ symmetry: one $\mathbb{Z}_3$ factor cyclically permutes $f_i$ while the other takes $f_i \rightarrow e^{\frac{2\pi i m_i}{3}} f_i$ with $m_1 = 0, m_2 = 1, m_3 = 2$.

At low doping, the Lagrangian $L[f_i,a]$ can be obtained by expanding the dispersion of each flavor near their minimum and including interactions between the flavors. Thus, we have
\begin{equation*}
    \sum_{i=1}^3 L[f_i, a] = \sum_{i=1}^3 \bar{f_i} \left(\partial_\tau - i a_0 - \frac{(\vec \nabla - i \vec a)^2}{2m} \right)f_i + L_{int}
\end{equation*} 
where $m$ is the CF effective mass. The term $L_{int}$ describes 4-fermion interactions that descend from the microscopic density-density interaction of the electrons. In Sec.~\ref{sec:SU(3)sym_NFL}, we argue that in the low doping limit, the dominant interactions may preserve an $SU(3)$ flavor rotation symmetry between the $f_i$.~\footnote{Strictly speaking this is an $SU(3)/Z_3$ symmetry as the $\mathbb{Z}_3$ center of $SU(3)$ can be absorbed by a gauge transformation involving $a$.} In the discussion below, we will first assume that the $SU(3)$-breaking interactions are strong and will return to the case of weak $SU(3)$-breaking in Sec.~\ref{sec:SU(3)sym_NFL}.

Away from $\nu = 1/3$, the physical dopant density is controlled by the flux of $\alpha$. Using the equations of motion for $\alpha$ and $a$, we see that
\begin{equation}
    \sum_{i=1}^3 \rho_{f_i} + \frac{1}{2\pi} \nabla \times (\bs{a} - \bs{\alpha}) = \frac{1}{2\pi} \nabla \times (2\bs{\alpha} + \bs{a}) = 0 \,.
\end{equation}
As a result, doping away from $\nu = 1/3$ forces a nonzero average magnetic field $\nabla \times \bs{a}$ such that each $f_i$ field is at a Landau level filling $\nu_{f_i} = -1/2$. 

Following Ref.~\cite{Shi2024_doping}, we remove the flux seen by $f_i$ through a \textit{secondary composite fermion} transformation. More precisely, we use the parton decomposition $f_i = \Psi_i \Phi_i$ such that the fermions $\Psi_i$ see no magnetic field and can hence form Fermi surfaces, while the bosons $\Phi_i$ see the emergent mean magnetic field and are at a Landau level filling $-1/2$ each. A natural option is to put each $\Phi_i$ is a bosonic Laughlin state $U(1)_{-2}$ at filling $\nu_{\Phi_i} = -1/2$. In Appendix~\ref{app:connect} we use the same decomposition $f_i = \hat \Psi_i \hat \Phi_i$ but choose a different topological state for $\hat \Phi_i$ at this filling to obtain the effective theory of Ref.~\cite{Fan2026_weakpairing_SC} (see Fig.~\ref{fig:Diff_sCF}). Which of these choices obtains is an energetic question beyond the scope of the parton/effective theory approach. 

Sticking for the time being with each of the $\Phi_i$ forming the bosonic Laughlin state, each sCF $\Psi_i$ sees zero magnetic field on average and can form a Fermi pocket. The full Lagrangian is then
\begin{equation}
    \begin{aligned}
    L_{\rm doped} &= \sum_{i=1}^3 \left\{L[\Psi_i, b_i] + \frac{2}{4\pi} \beta_i d \beta_i - \frac{1}{2\pi} \beta_i d (a - b_i)\right\}  \\
    &\hspace{1cm} + \mathrm{CS}[a,g] - \frac{2}{4\pi} \alpha d \alpha + \frac{1}{2\pi} \alpha d (A-a) \,,
    \end{aligned}
\end{equation}
where $b_i$/$\beta_i$ are dynamical $\mathrm{spin}_{\mathbb{C}}$ connections/$U(1)$ gauge fields introduced by the sCF transformation. In Appendix~\ref{app:dope_simplified}, we show that many of the relevant gauge fields can be integrated out, leaving us with a much simpler Lagrangian (at the cost of not keeping the $\mathbb{Z}_3$ symmetry exchanging the three flavors manifest) %\ts{One disadvantage of this formulation is that it makes the $SU(3)$  or even the $\mathbb{Z}_3$ symmetry non-manifest and hence the IR realization of translation symmetry.}
\begin{equation}\label{eq:doped_simple}
    \begin{aligned}
    L_{\rm doped} &= L[\Psi_1, - A - 2 \beta] + L[\Psi_2, b] + L[\Psi_3, 2 \beta - b] \\
    &\hspace{1cm} - \mathrm{CS}[b,g] + \frac{1}{2\pi} \beta d (b - A - 2 \beta) \,. 
    \end{aligned}
\end{equation}
%As a sanity check, note that when $\Psi_i$ are gapped, integrating out $b$ generates the standard TQFT for $U(1)_3$, which describes the Laughlin state at $\nu = 1/3$. 

%\ts{Also some of the above can be shortened as it is described in some detail in our PRX paper.}

\section{Gauge-field-mediated pairing instability} 

%At nonzero doping away from $\nu = 1/3$, the $\Psi_i$ fermions see zero average magnetic field and form degenerate Fermi pockets. 
At intermediate temperatures, the $\Psi_i$ Fermi surfaces strongly interact with the emergent gauge fields $b$ and $\beta$, giving rise to a non-Fermi liquid metal dubbed the \textit{secondary composite Fermi liquid} (sCFL) in Ref.~\cite{Shi2024_doping}. However, this sCFL state is unstable to pairing at sufficiently low temperature. Through a one-loop renormalization group analysis, Ref.~\cite{Shi2024_doping} showed that the three fermion flavors always have an inter-flavor pairing instability mediated by the dynamical gauge fields (the same conclusion was reached through a different argument in Ref.~\cite{Fan2026_weakpairing_SC}). More precisely, let $V_{ij}$ be the BCS pairing interaction between flavor $i$ and $j$. Under renormalization group flow, the intra-flavor attraction $V_{ii}$ flows to a finite fixed point value while the inter-flavor attraction $V_{i \neq j}$ has a runaway flow independent of initial conditions. Moreover, due to the $\mathbb{Z}_3$ symmetry between the three flavors enforced by lattice translation symmetry, $V_{12}, V_{13}, V_{23}$ have identical flow equations~\cite{Shi2024_doping}. Through a simple Ginzburg-Landau analysis in Appendix~\ref{app:LG}, we find that the possible zero-temperature paired states fall into two categories, depending on effective interactions between the pair fields $\Delta_{ij} \sim \Psi_i \Psi_j$ that go beyond the one-loop RG analysis. 

The first category of phases arises when the $\mathbb{Z}_3$ flavor symmetry is spontaneously broken. In Appendix~\ref{app:LG}, we show that within a quartic Landau-Ginzburg theory, the partially polarized state with $\Delta_{12} \neq \Delta_{13} \neq \Delta_{23} \neq 0$ is never energetically favored. Therefore, we can focus on the fully polarized case where two of the three pockets pair up in an angular momentum channel $l$, while the third pocket remains unpaired.\footnote{Due to the $\mathbb{Z}_3$ symmetry, the final electronic phase does not depend on which two pockets we choose to pair up.} The universal effective theory does not uniquely fix the pairing channel. Borrowing numerical results from bilayer CFLs at $\nu_{\rm tot} = -1/2 - 1/2$~\cite{Milovanovic2007_CFpairing,Moller2008_CFpairing,Moller2009_CFpairing,Milovanovic2015_CFpairing}, we expect the pairing angular momentum to be $l = 1$ corresponding to $p + ip$. Remarkably, the analysis of Ref.~\cite{Shi2024_doping} shows that the resulting phase of matter is an ordinary Fermi liquid coexisting with a period-3 charge density wave, with all dynamical gauge fields completely Higgsed. This state is the CDW metal in Table~\ref{tab:scnflnear2/3}.

In the second category, the mean-field pairing order parameter matrix preserves the $\mathbb{Z}_3$ flavor symmetry
\begin{equation}\label{eq:OP_matrix}
    \hat \Delta = \Delta e^{i l \theta_{\bs{k}}} \begin{pmatrix}
        0 & 1 & 1 \\ 1 & 0 & 1 \\ 1 & 1 & 0 
    \end{pmatrix}  \,. 
\end{equation}
This kind of flavor-symmetric pairing was recently proposed by Ref.~\cite{Fan2026_weakpairing_SC} under a different effective field theory and was not considered in our earlier work~\cite{Shi2024_doping}. 

We emphasize that although the pairing instability is a universal prediction of the sCFL theory, which of these two pairing patterns gets realized is determined by details of the microscopic interactions and is outside the scope of any effective field theory. 

In what follows, we will consider the flavor-symmetric pairing channel and derive the corresponding electronic phase in the strong and weak pairing limits. As we will see, in contrast to the chiral topological superconductors found in Ref.~\cite{Fan2026_weakpairing_SC}, the same flavor-symmetric pairing applied to our theory~\eqref{eq:doped_simple} gives SC* states, in which a charge-$2e$ chiral topological superconductor coexists with a $\mathbb{Z}_2$ topological order. That such an exotic superconducting state occurs quite naturally within a standard CF treatment of the fluid of  $1/3$ charged anyons obtained by doping FQAH/FCI states raises the prospects for its realization in experiments.

\subsection{Strong pairing limit}

In the strong pairing limit, the effective field theory for the flavor-symmetric paired state involves three dynamical $U(1)$ gauge fields $\lambda_{12}, \lambda_{13}, \lambda_{23}$
\begin{equation}
    \sum_{i=1}^3 L[\Psi_i, A_i] \rightarrow \sum_{i < j \leq 3} \frac{1}{2\pi} \lambda_{ij} d (A_i + A_j)  \,. 
\end{equation}
When applied to our setup, we have the identification
\begin{equation}
    A_1 = -A - 2 \beta \,, \quad A_2 = b \,, \quad A_3 = 2 \beta - b \,. 
\end{equation}
From this identification, we see that integrating over $\lambda_{23}, \lambda_{12}$ sets $2 \beta = A - b = 0$. Plugging these constraints into the full Lagrangian in \eqref{eq:doped_simple} gives 
\begin{equation*}
    L_{\rm strong} = \frac{2}{2\pi} \lambda_{13} d A + \frac{2}{2\pi} \lambda_{23} d \beta - \mathrm{CS}[A,g] - \frac{4}{2\pi} \beta d \beta \,. 
\end{equation*}
Shifting $\lambda_{23} \rightarrow \lambda_{23} + \beta$, we find the final strong pairing superconductor
\begin{equation}\label{eq:strong_pairing}
    L_{\rm strong} = \frac{2}{2\pi} \lambda_{13} d A - \mathrm{CS}[A,g] + \frac{2}{2\pi} \lambda_{23} d \beta \,. 
\end{equation}
The first two terms describe a charge-$2e$ chiral superconductor with chiral central charge $c_- = -1$. The final term describes a residual $\mathbb{Z}_2$ topological order that coexists with the condensate. 

\subsection{Weak pairing limit}

Next, we turn to the weak pairing limit. Diagonalizing the pairing matrix~\eqref{eq:OP_matrix}, we find three decoupled Bogoliubov branches with eigenvalues $(\Delta, \Delta, - 2 \Delta) e^{il \theta_{\bs{k}}}$. Choosing $l = -1$ as before, the superconductor of $\Psi_i$ contributes a chiral central charge $c_- = -3/2$. Following Ref.~\cite{Ma2020_QCDtransition,Shi2025_dopeMR}, we can describe this state by a pair of $U(1)$ gauge fields $\lambda_{23}, \lambda_{13}$ and a $U(2)$ gauge field $\gamma$
\begin{equation*}
    \begin{aligned}
    &\sum_{i=1}^3 L[\Psi_i, A_i] \rightarrow \frac{1}{2\pi} \lambda_{12} d (A_1 + A_2) + \frac{1}{2\pi} \lambda_{23} d (A_2 + A_3) \\
    &+ \frac{2}{4\pi} \Tr \left[\gamma d \gamma + \frac{2}{3} \gamma^3\right] - \frac{1}{4\pi} \Tr \gamma \, d \Tr \gamma + \frac{1}{2\pi} \Tr \gamma \, d A_2 \,.
    \end{aligned}
\end{equation*}
The $U(1)$ part of $\gamma$ implements the constraint $2A_2 = 0$. Combined with the constraints imposed by $\lambda_{12}, \lambda_{23}$, we arrive at the correct $U(1)^3$ symmetry-breaking pattern $A_1 + A_2 = A_2 + A_3 = A_1 + A_3 = 0$. The nontrivial half-integer chiral central charge $c_- = -3/2$ is captured by the $SU(2)_{-2}$ sector of the $U(2)_{-2,0}$ gauge theory of $\gamma$. 

When applied to our setup, we find
\begin{equation*}
    \begin{aligned}
    L_{\rm weak} &= \frac{2}{4\pi} \Tr \left[\gamma d \gamma + \frac{2}{3} \gamma^3\right] - \frac{1}{4\pi} \Tr \gamma \, d \Tr \gamma\\
    & + \frac{1}{2\pi} \Tr \gamma \, d A - \frac{4}{4\pi} \beta d \beta + \frac{2}{2\pi} \lambda_{23} d \beta - \mathrm{CS}[A,g] \,. 
    \end{aligned}
\end{equation*}
Shifting $\lambda_{23} \rightarrow \lambda_{23} + \beta$ gives a clean final form
\begin{equation}\label{eq:weak_pairing}
    \begin{aligned}
    L_{\rm weak} &= \frac{2}{4\pi} \Tr \left[\gamma d \gamma + \frac{2}{3} \gamma^3\right] - \frac{1}{4\pi} \Tr \gamma \, d \Tr \gamma\\
    &\hspace{0.6cm}- \mathrm{CS}[A,g] + \frac{1}{2\pi} \Tr \gamma \, d A + \frac{2}{2\pi} \lambda_{23} d \beta \,. 
    \end{aligned}
\end{equation}
The first four terms now describe a topological superconductor with $c_- = -5/2$, while the last term describes a stacked $\mathbb{Z}_2$ topological order. 

Compared to the strong pairing case, we see that the intrinsic topological order remains the same, while the chiral central charge shifts by $-3/2$. There is a very simple physical understanding of this result. Before pairing, the fermion fields $\Psi_i$ are coupled to $-A - 2 \beta, b, 2\beta - b$. When the pairing condensate forms, all of these gauge fields get identified with $A$. As a result, each $\Psi_i$ gets identified with a physical Bogoliubov fermion in the final superconductor. The transition from weak to strong pairing of $\Psi_i$ therefore translates to a transition from weak to strong pairing of the physical Bogoliubov fermions. This explains why the intrinsic topological order is unchanged across the transition, despite a jump $\Delta c_- = -3/2$. 

Under a particle-hole transformation, the strong and weak pairing SC* states derived above retain the toric code topological order but have modified chiral central charges $c_- = 1 - (-1) = 2$ and $c_- = 1 - (-5/2) = 7/2$. These are the values advertised in the introduction for the doped $\nu = 2/3$ Jain state. 

\section{Fate of anyons across the FCI-SC transition}
 
As emphasized both in Ref.~\cite{Shi2024_doping} and Ref.~\cite{Fan2026_weakpairing_SC}, pairing of fermionic partons provides a mechanism for superconductivity near an FQAH that is conceptually distinct from previous routes~\cite{Laughlin1988_anyonSC,Lee1989_anyonSC,Fetter1989_anyonSC_RPA,Chen1989_anyonSC,Wen1990_anyonSC,Tang2013_anyonSC,Kim2024_anyonSC,Shi2024_doping,Divic2024_HofHubb,Zhang2025_SU(3)1dope,Pichler2025_anyonSC,Shi2025_dopeMR,Nosov2025_plateau,Shi2025_anyon_delocalization,Han2025_anyonexciton,Nakajima2025_thermo_anyon,Kuhlenkamp2025_HofHubb,Shi2025_nonAbelian_TSC}. 
% Crucially, earlier effective field theories introduced doped anyons through a flux-attached fermion/boson field that sees \textit{nonzero emergent magnetic field} upon doping away from the parent FCI. In this framework, superconductivity emerges when these fermion/boson fields form gapped quantum Hall states consistent with their average filling fraction. However, the generalized CFs/secondary CFs considered in Ref.~\cite{Fan2026_weakpairing_SC}/this paper see \textit{zero magnetic field on average} and form superconducting states.

This conceptual distinction leads to a different phenomenology for the chemical potential tuned FCI-SC transition in the presence of quenched disorder.  For a class of theories ({\it e.g.} superconductors obtained by doping the $2/3$ anyons), it was argued that localized anyons evolve into localized vortices across the FCI-SC transition~\cite{Shi2025_anyon_delocalization,Shi2025_dopeMR}. As a result, the superconductor close to the phase transition hosts a random sprinkling of vortices and anti-vortices with vanishing net vorticity.\footnote{The preferred distribution of vorticities near the FCI-SC transition are determined by the parent FCI and generally different from the distribution of vorticities in the spontaneous vortex-antivortex lattice state found in Ref.~\cite{Gaggioli2025_vortex_antivortex}.} This unusual SC was dubbed an anomalous vortex glass (AVG)~\cite{fisher1989vortex,fisher1991thermal,fisher1991vortex,Shi2025_anyon_delocalization}. However, in mechanisms based on pairing fermionic partons, the onset of $\Delta_{i \neq j}$ Higgses the dynamical gauge fields such that $\Psi_i$ only carries charge under the unbroken $\mathbb{Z}_2^f$ background field $A$. An immediate implication is that a localized charge $1/3$ anyon in the Laughlin state sourced by $\Psi_i$ evolves into a localized Bogoliubov quasiparticle in the SC phase. In the vicinity of the phase transition, we therefore expect to find an ordinary charge-$2e$ SC with disorder-pinned Bogoliubov quasiparticles rather than a more exotic AVG phase. 

\section{Approximate SU(3) symmetry and a possible non-Fermi liquid metal}\label{sec:SU(3)sym_NFL}
We now turn to a more careful consideration of the possibility of approximate $SU(3)$ global symmetry for the doped CFs in the low density limit. A reasonable assumption is that the dominant four-fermion interactions between the CFs $f_i$ descend from the microscopic electronic density-density interaction: 
\begin{equation} 
 \int_q  V(|\bs{q}|) \rho(\bs{q}) \rho(-\bs{q}) 
\end{equation} 
where $V$ is the Fourier transform of the interaction potential ({\it e.g.} a screened Coulomb interaction) and $\rho(\bs{q})$ is the Fourier transform of the electron density at wave vector $\bs{q}$. We expect that $V(|\bs{q}|)$ is a decreasing function of $|\bs{q}|$. In the usual construction, the CF density is the same as the electron density. Thus, in the low energy limit we have 
\begin{equation} 
    \rho(\bs{r}) \approx \sum_i f^\dagger_i (\bs{r}) f_i(\bs{r}) + \sum_{i \neq j} e^{i \bs{Q}_{ij} \cdot \bs{r}} f^\dagger_i (\bs{r}) f_j (\bs{r}) 
\end{equation} 
where the vectors $\bs{Q}_{ij}$ denote the vectors connecting the different valleys and have magnitude of order the inverse microscopic lattice constant.  
The resulting CF interaction Lagrangian is then 
\begin{equation} 
\label{eq: CFint} 
    \begin{aligned}
    L_{int} &\approx \int_{\bs{q}}  \left[V_0 \sum_i \rho_i (\bs{q}) \sum_j \rho_j (- \bs{q}) - V_1 \sum_{i \neq j} \rho_i (\bs{q}) \rho_j(-\bs{q}) \right]
    \end{aligned}
\end{equation} 
where $V_0$/$V_1$ is the component of $V(|\bs{q}|)$ at $\bs{q}=0$/$\bs{q} = \pm \bs{Q}_{12}$ and the momentum integration is restricted to $|\bs{q}| \ll |\bs{Q}_{ij}|$. The first term preserves $SU(3)$ flavor symmetry while the second term breaks it. Thus, if the first term dominates, the doped CF theory has approximate $SU(3)$ symmetry. 

The secondary CF construction given above (as well as the different one in Appendix~\ref{app:connect} that yields the theory of Ref. \cite{Fan2026_weakpairing_SC}) strongly breaks this $SU(3)$ symmetry. Thus, if $SU(3)$ holds approximately, we must seek a different $SU(3)$-symmetric construction. A natural choice is to use a different parton decomposition $f_i = \chi_i \Phi$ where the fermionic partons $\chi_i$ transform in the fundamental of $SU(3)$ and $\Phi$ is an $SU(3)$ singlet boson. As the $f_i$ are at a total Landau level filling of $-3/2$, we put $\Phi$ in a bosonic Jain state at that filling while the $\chi_i$ see zero magnetic field on average and form a Fermi surface. The resulting state is analysed in Appendix~\ref{app: z3om}. Remarkably it reduces to a metal of 3 fermion species, each with electric charge $1/3$ coupled to a ``twisted" \footnote{The ``twisting" here means that this belongs to the Dijkgraaf-Witten family of $Z_3$ gauge theories as is clear from the Lagrangian in Eqn.~\ref{eq: z3om}.} discrete $\mathbb{Z}_3$ gauge field. The electron has thus fractionalized into three charge-$1/3$ fermions in this metal.  
Due to the similarity with Orthogonal Metals described in Ref.~\cite{nandkishore2012orthogonal}, we will refer to this phase as a $\mathbb{Z}_3$ Orthogonal Metal (Z3OM). 

The effective Lagrangian derived in Appendix \ref{app: z3om} for this state is 
\begin{equation}
\label{eq: z3om}
 \sum_{i=1}^3 L[\chi_i, b] + \frac{1}{2\pi} \alpha d (A - 3 b) + 3 \mathrm{CS}[b,g]   
\end{equation}
where the $\chi_i$ form Fermi surfaces, and $b$ is a spin$_\mathbb{C}$ connection. The reduction to a $\mathbb{Z}_3$ gauge field is due to the coupling to the $U(1)$ gauge field $\alpha$. We see that the last term gives a contribution to the Hall conductivity $\sigma_{xy} = 1/3$. However, the total Hall conductivity is not quantized due to the presence of metallic quasiparticles $\chi_i$. The discreteness of the gauge field means that the $\chi_i$ fields describe well-defined quasiparticles at low energy near their Fermi surfaces which carry electric charge $1/3$. 

A different perspective on this low energy theory is obtained by defining a new spin$_\mathbb{C}$ connection $a$ through $b =  a + \alpha$ so that the Lagrangian becomes 
\begin{equation} 
\sum_{i=1}^3 L[\chi_i, a + \alpha ] + \frac{1}{2\pi} \alpha dA  - \frac{3}{4\pi}\alpha d\alpha +  3 \mathrm{CS}[a,g] \,. 
\end{equation} 
Let us consider the theory defined by the Chern-Simons terms alone. The $\alpha$ sector describes the usual $1/3$ Laughlin state. The electrically neutral $a$-sector has $3$ distinct anyons with statistical angles $0, \pi - \frac{\pi}{3}, -\frac{4\pi}{3}$. The shift by $\pi$ for one of the anyons (as compared to the usual $U(1)_{-3}$) is because $a$ is a spin$_\mathbb{C}$ connection. It follows that an anyon that has charge $+1$ under both $\alpha$ and $a$ is a fermion that carries physical $U(1)_A$ charge $1/3$. The structure of this effective Lagrangian is reminiscent of those obtained in Ref.~\cite{Shi2025_dopeMR} in the ``stack and condense" route to understanding anyon-driven superconductivity. Here stacking the Laughlin state with the neutral anyon theory (the $a$-sector) enables the formation of electrically charged fermions which can then form a Fermi sea. 
 
The large Hall conductivity of the Z3OM state  is accompanied by a longitudinal conductivity proportional to the dopant density. Its sharp Fermi surfaces will lead to quantum oscillations but there will be no sharp electron quasiparticle in tunneling or in the momentum-resolved electron spectral function. Shot noise measurements will reveal the fractional $1/3$ charge of the fermionic charge carriers. Thus we have the very interesting possibility of an experimental realization of what is possibly the simplest non-Fermi liquid metal in $2d$ when the $2/3$ or $1/3$ FCIs are doped with charge-$1/3$ anyons. It will be fascinating to look for evidence of this state in current and future experimental platforms.

\section{Discussion}

%In this paper we continued our exploration of the possible ground states of the mobile anyon fluid induced by doping FQAH/FCI states. Our focus was on a fluid of $1/3$ charged anyons near the $1/3$ or $2/3$ FCI. We described these anyons within the standard CF framework (which is expected to capture the microscopic physics). The CFs feel an emergent mean magnetic field at low doping which leads to the formation of CF Landau levels. In our previous work, we employed a secondary CF construction to describe some possible ground states that result from a pairing instability of their Fermi surfaces (which included a charge ordered Fermi liquid metal as well as an exotic Pair Density Wave superconductor). Here we showed how a different pairing channel -  inspired by  Ref. \cite{Fan2026_weakpairing_SC} - leads very naturally to a chiral topological superconductor coexisting with ``toric code" topological order. 
%We also show, in Appendix~\ref{app:connect}, how a different construction of secondary CFs leads to the theory of Ref. \cite{Fan2026_weakpairing_SC}. A summary of the superconducting states described so far in this $1/3$ anyon fluid is in Table. 

%If the residual CF interactions have an approximate $SU(3)$ valley symmetry, then the states described previously are not appropriate. Instead we described a simple but fascinating non-fermi liquid metal where there are 3 species of charge-$1/3$ fermions that form fermi surfaces, and are coupled to a $Z_3$ gauge field. 

The results in this paper, as well as Refs.~\cite{Shi2024_doping, Shi2025_dopeMR,Zhang2025_SU(3)1dope,Shi2025_nonAbelian_TSC,Fan2026_weakpairing_SC, Wang2026_U3anyonSC}, show that there is room for much interesting physics in a mobile fluid of charge-1/3 anyons with statistics $\pm \pi/3$. Discerning between the different itinerant phases (see Table~\ref{tab:scnflnear2/3}) will require a more detailed description of the different states as realized in microscopic models of the experimental systems. Below we summarize our current understanding. 

\subsubsection{Comments on energetics}
In both tMoTe$_2$ and the multilayer rhombohedral graphene platform, several Jain states as well as the Composite Fermi Liquid are seen. This experimental input suggests~\footnote{In Ref.~\cite{Shi2025_anyon_delocalization} we proposed that the observation of Jain states alone does not inform whether standard CFs capture the doping from $2/3$ filling {\it away} from $1/2$, as there are no Jain states between $2/3$ and $1$. It remains to be seen if this is true. Experimentally~\cite{Xu2025_SCdopeTMD} there is a clear asymmetry between the doped states on either side of $2/3$.} that upon doping from $2/3$ filling toward $1/2$, the standard CF formalism is likely an excellent starting point in describing the energetics of existing experimental realizations of FCI states. Upon doping away from $2/3$, we know that the three flavors of CFs see an effective magnetic field such that they each are at total effective Landau level filling $3/2$. The main question is then: what state do the CFs form at this filling? Clearly, the answer to this question depends on the residual interactions between the CF flavors (see Fig.~\ref{fig:Diff_sCF} for several competing possibilities discussed throughout the paper). We argued that in the limit $V_1 \ll V_0$ in Eq.~\eqref{eq: CFint}, the approximate $SU(3)$ symmetry is best described by writing the CFs $f_i = \chi_i \Phi$ which leads to the Non-Fermi Liquid Z3OM state.  

What if  $V_1$ is only slightly smaller than, or even comparable to $V_0$? To see what may happen, it is useful to rewrite Eq.~\eqref{eq: CFint} as 
\begin{eqnarray*} 
    L_{int} &\approx \int_{\bs{q}}  \left( V_0 - V_1 \right) \sum_i \rho_i (\bs{q}) \sum_j \rho_j (- \bs{q}) \\
    & + V_1 \sum_{i} \rho_i (\bs{q}) \rho_i(-\bs{q})  \,.
\end{eqnarray*} 
If $V_1 > V_0/2$, the total charge density repulsion is smaller than the intra-flavor repulsion. If the intra-flavor repulsion dominates, then we expect each $f_i$ to form its own sCFL state that is captured by the construction $f_i = \Psi_i \Phi_i$ in which each $\Phi_i$ forms its own bosonic Laughlin state. Within this construction, the sCF pairing instability leads to the novel SC$^*_{7/2}$ superconductor in this paper, or to the CDW metal in Ref.~\cite{Shi2024_doping}. In principle, the even more exotic PDW SC* arising from intra-flavor pairing is also allowed, though it is energetically less likely than inter-flavor pairing~\cite{Shi2024_doping,Fan2026_weakpairing_SC}. 

What about the $\widehat{s\mathrm{CFL}}$ state? As described in Appendix~\ref{app:connect}, this state requires writing the CFs as $f_i = \hat\Psi_i \hat \Phi_i $ and putting $\hat\Phi_i$ into a topological phase described by a $3\times 3$ K-matrix with zero diagonal elements (c.f. Fig.~\ref{fig:Diff_sCF}). The corresponding Landau level wavefunction has zeroes when two CFs of different flavors approach each other but no zero when two CFs of the same flavor approach each other. Thus we might guess that this state is not energetically preferred if the interaction has the structure described in Eq.~\eqref{eq: CFint}. This argument suggests that the theory based on $\widehat{s\mathrm{CF}}$ $\hat{\Psi}_i$ - and the resulting superconductors found in Ref.~\cite{Fan2026_weakpairing_SC} - might not be energetically competitive as long as standard CFs are a good starting point. 

\subsubsection{Numerics}

Existing numerical works~\cite{Guerci2025_FCISC,Wang2025_SCmeltFCI,guerci2026topological} mostly focus on a bandwidth-tuned transition at fixed filling from the $\nu = 2/3$ Jain FCI to a chiral topological superconductor with $c_- = -1/2$.   In the adiabatic limit of the skyrmion crystal model for twisted MoTe$_2$, Refs.~\cite{Guerci2025_FCISC,guerci2026topological} provide evidence that this transition is strongly first order, and assert that it has nothing to do with anyon-driven superconductivity. In a simpler toy model with fractionally filled Landau level + tunable periodic potential~\cite{Wang2025_SCmeltFCI}, the nature of the bandwidth-tuned transition is not settled and it is not currently clear how to understand the SC that is obtained as a melted FCI state at fixed electronic filling. For the doped system, Ref.~\cite{guerci2026topological} provided evidence for SC$_{-1/2}$  at relatively large dopings ($\nu = 0.74$ and above) in a parameter regime where at $2/3$ the ground state is an FCI. The large doping makes it unclear whether this state could be understood as emerging from the FCI through an anyonic mechanism for superconductivity. 

\subsubsection{Experiments} Experimentally, superconductivity has been reported~\cite{Xu2025_SCdopeTMD} near the 2/3 FQAH state in tMoTe$_2$, and has been interpreted by us in Ref.~\cite{Shi2025_anyon_delocalization}, as well as in Ref.~\cite{Nosov2025_plateau}, in terms of anyon-driven superconductivity. Accepting that {\it some} anyonic mechanism is at play, we can ask which of the various SCs summarized in Table~\ref{tab:scnflnear2/3} is actually realized. Unfortunately there is currently not enough information to answer this question with confidence. If the SC has half-integer $c_-$, then it can also be understood as a weak coupling instability of the ordinary Fermi liquid  metal that is known to exist for $\nu \gtrsim 0.75$. However, the chemical potential also crosses a van Hove singularity around $\nu = 0.75$, and it is possible that there is a first order transition in the normal state that separates the superconducting range of fillings from those beyond $\nu \approx 0.75$. Thus the possibility of a SC with $c_- = -2$ or even one of the SC$^*$ phases in Table~\ref{tab:scnflnear2/3} are not currently ruled out. 

\subsubsection{Beyond superconductivity}

Going beyond superconductivity, it will also be interesting to look for the SU(3)-symmetric Z3OM in the vicinity of the $2/3$ or $1/3$ FCI. This non-Fermi liquid metal has roughly the same Hall conductivity as the parent FCI but has a metallic $\rho_{xx}$, despite the absence of sharp peaks in the momentum-resolved electron spectral function. Perhaps this state appears in a range of fillings between the FCI plateau and one of the candidate SCs in Table~\ref{tab:scnflnear2/3}, where violations of the approximate SU(3) symmetry (rotating the three translation-related CF flavors) are expected to be weak. \newline 

\textit{Note added:} since the arXiv posting of our paper, several related works have appeared on the arXiv.  Refs.~\cite{Senthil2026_Z3OM,Han2026_Z3OM} construct a distinct $Z_3$ Orthogonal Metal with a single Fermi surface of charge-$1/3$ fermions, and elaborate on their phenomenology and experimental signatures. Further discussions of anyon-driven superconductors in the charge-$1/3$ anyon fluid have also appeared in Refs.~\cite{Zhang2026_colorSC,Mehta2026_colorSC}. 

\section*{Acknowledgment}

We would like to thank Zhaoyu Han and Ashvin Vishwanath for discussions. ZDS was supported by a Leinweber Institute for Theoretical Physics postdoctoral fellowship at Stanford University and in part by the Gordon and Betty Moore Foundation EPiQS initiative, Grant GBMF8686.01. T.S. was supported by the U.S. Department of Energy under Grant DE-SC0008739.

\bibliography{DopeCF}% Produces the bibliography via BibTeX.

\onecolumngrid
\newpage 
\appendix 

\section{Connecting with the effective field theory in Ref.~\cite{Fan2026_weakpairing_SC}}
\label{app:connect}

In this Appendix, we explain how the standard CF construction used in this paper can be connected to the effective field theory in Ref.~\cite{Fan2026_weakpairing_SC}. 

We start with the standard CF Lagrangian derived in the main text, written in terms of continuum fields $f_i$ that create excitations near three degenerate valleys in momentum space
\begin{equation}
    L_{\rm doped} = \sum_{i=1}^3 L[f_i, a] + \mathrm{CS}[a,g] - \frac{2}{4\pi} \alpha d \alpha + \frac{1}{2\pi} \alpha d (A-a) \,. 
\end{equation}
At nonzero doping away from $\nu = 1/3$, each fermion field $f_i$ is at Landau level filling $\nu_{f_i} = -1/2$ relative to the emergent magnetic field $\nabla \times \bs{a}$ it sees. This filling fraction motivated a secondary CF transformation in which we bind each $f_i$ with two flux quanta of $a$ to build secondary CFs $\Psi_i$ that see zero magnetic field. On a lattice, this transformation is implemented by further decomposing each $f_i$ as $f_i = \Psi_i \Phi_i$ and choosing a mean-field in which each hard-core boson $\Phi_i$ sees the total magnetic field of $a$, while $\Psi_i$ sees zero magnetic field on average. Putting each $\Phi_i$ in a $U(1)_{-2}$ state consistent with the $\nu_{\Phi_i} = -1/2$ filling fraction then reproduces the desired flux attachment
\begin{equation}
    \sum_{i=1}^3 L[f_i, a] \rightarrow \sum_{i=1}^3 \left\{L[\Psi_i, b_i] + \frac{2}{4\pi} \beta_i d \beta_i - \frac{1}{2\pi} \beta_i d (a - b_i)\right\} \,. 
\end{equation}

While this flux attachment procedure is natural when the three flavors $f_1, f_2, f_3$ are decoupled, other options become possible when we include interactions between the flavors. To reproduce the theory in Ref.~\cite{Fan2026_weakpairing_SC}, we use the same decomposition $f_i = \hat \Psi_i \hat \Phi_i$ but choose a different ansatz for the three bosonic partons $\hat \Phi_i$ at total Landau level filling $\nu_{\hat \Phi} = - 3/2$ described by a $3 \times 3$ K-matrix
\begin{equation}
    \sum_{i=1}^3 L[\hat \Phi_i, a - b_i] \rightarrow - \frac{1}{4\pi} \alpha^T K d \alpha + \sum_{i=1}^3 \frac{1}{2\pi} \alpha_i d (a - b_i) \,, \quad K = - \begin{pmatrix}
        0 & 1 & 1 \\ 1 & 0 & 1 \\ 1 & 1 & 0
    \end{pmatrix} \,. 
\end{equation}
One can verify that this K-matrix theory has a Hall conductance $\sigma_{xy} = -3/2$ consistent with the boson filling fraction $\nu_{\phi} = -3/2$ and has $|\det K| = 2$. Moreover, since the K-matrix preserves the $\mathbb{Z}_3$ symmetry (the IR image of a subgroup of UV lattice translations) that cyclically permutes the boson flavors $\hat \Phi_i$, this theory also preserves lattice translation symmetry. By the arguments in Ref.~\cite{Cheng2025_orderingqh}, this topological order has the minimal anyon count consistent with the Landau level filling of $\{\hat \Phi_i\}$.

Plugging this bosonic sector into the full Lagrangian, we find 
\begin{equation}
    L_{\rm doped} = \sum_{i=1}^3 L[\hat \Psi_i, b_i] + \frac{1}{2\pi} (\alpha_1 d \alpha_2 + \alpha_1 d \alpha_3 + \alpha_2 d \alpha_3) + \sum_{i=1}^3 \frac{1}{2\pi} \alpha_i d (a - b_i) + \mathrm{CS}[a,g] - \frac{2}{4\pi} \alpha d \alpha + \frac{1}{2\pi} \alpha d (A-a) 
\end{equation}
Integrating over $a$ gives
\begin{equation}
    \begin{aligned}
    L_{\rm doped} &= \sum_{i=1}^3 L[\hat \Psi_i, b_i] + \frac{1}{2\pi} (\alpha_1 d \alpha_2 + \alpha_1 d \alpha_3 + \alpha_2 d \alpha_3) - \sum_{i=1}^3 \frac{1}{2\pi} \alpha_i d b_i \\
    &\hspace{0.5cm} - \frac{1}{4\pi} (\sum_{i=1}^3 \alpha_i - \alpha) d (\sum_{i=1}^3 \alpha_i - \alpha) - \frac{2}{4\pi} \alpha d \alpha + \frac{1}{2\pi} \alpha d A \\
    &= \sum_{i=1}^3 L[\hat \Psi_i, b_i] - \frac{1}{4\pi} \sum_{i=1}^3 \alpha_i d \alpha_i - \sum_{i=1}^3 \frac{1}{2\pi} \alpha_i d (b_i - \alpha) - \frac{3}{4\pi} \alpha d \alpha + \frac{1}{2\pi} \alpha d A 
    \end{aligned}
\end{equation}
Each $\alpha_i$ now has a unit Chern-Simons level and can be integrated out. The resulting theory is
\begin{equation}
    L_{\rm doped} = \sum_{i=1}^3 L[\hat \Psi_i, b_i] + \sum_{i=1}^3 \mathrm{CS}[b_i, g] + \frac{1}{2\pi} \alpha d (A - b_1 - b_2 - b_3) \,. 
\end{equation}
Integrating over $\alpha$ imposes a constraint $b_3 = A - b_1 - b_2$. Subsequently integrating out $b_3$ gives
\begin{equation}\label{eq:match_Fan_1/3}
    \begin{aligned}
    L_{\rm doped, 1/3} &= L[\hat \Psi_1, b_1] + L[\hat \Psi_2, b_2] + L[\hat \Psi_3, A - b_1 - b_2] + \frac{1}{4\pi} b_1 d b_1 + \frac{1}{4\pi} b_2 d b_2 + \frac{1}{4\pi} (A - b_1 - b_2) d (A - b_1 - b_2) + 6 \mathrm{CS}_g \\
    &= L[\hat \Psi_1, b_1] + L[\hat \Psi_2, b_2] + L[\hat \Psi_3, A - b_1 - b_2] + \frac{1}{4\pi} b^T \tilde K b + 4 \mathrm{CS}_g - \frac{1}{2\pi} (b_1 + b_2) d A + \mathrm{CS}[A,g] \,,
    \end{aligned}
\end{equation}
where the final K-matrix is given by $\tilde K = \begin{pmatrix} 2 & 1 \\ 1 & 2 \end{pmatrix}$. 

The Lagrangian in \eqref{eq:match_Fan_1/3} was derived from doping the Laughlin state at $\nu = 1/3$. To obtain the doped theory near $\nu = 2/3$, we can simply perform a particle-hole transformation, which reverses the signs of all Chern-Simons terms and adds an extra $\mathrm{CS}[A,g]$. The resulting theory near $\nu = 2/3$ is 
\begin{equation}\label{eq:match_Fan_2/3}
    \begin{aligned}
    L_{\rm doped, 2/3} = L[\hat \Psi_1, b_1] + L[\hat \Psi_2, b_2] + L[\hat \Psi_3, A - b_1 - b_2] - \frac{1}{4\pi} b^T \tilde K b - 4 \mathrm{CS}_g + \frac{1}{2\pi} (b_1 + b_2) d A \,.
    \end{aligned}
\end{equation}
This final Lagrangian precisely matches Eq.~(23) in Ref.~\cite{Fan2026_weakpairing_SC}. 

In summary, we find that the fermions $\psi_i$ in Ref.~\cite{Fan2026_weakpairing_SC} can be identified with $\hat \Psi_i$, which is derived from the standard CF effective theory by fractionalizing doped CF fields $f_i$ as $f_i = \hat \Psi_i \hat \Phi_i$ and choosing the minimal flavor-symmetric ansatz for $\{\hat \Phi_i\}$ consistent with its total Landau level filling $\sum_{i=1}^3 \nu_{\hat \Phi_i} = -3/2$. In contrast, the secondary CFs $\Psi_i$ introduced in Ref.~\cite{Shi2024_doping} and analyzed in this paper corresponds to the ansatz in which each $\Phi_i$ forms a gapped state consistent with its Landau level filling $\nu_{\Phi_i} = -1/2$. 

\section{Simplified presentation of the doped theory}\label{app:dope_simplified}

Let us recall the Lagrangian for the secondary CFL, as written in the main text
\begin{equation}
    L = \sum_{i=1}^3 \left\{L[\Psi_i, b_i] + \frac{2}{4\pi} \beta_i d \beta_i - \frac{1}{2\pi} \beta_i d (a - b_i)\right\} + \mathrm{CS}[a,g] - \frac{2}{4\pi} \alpha d \alpha + \frac{1}{2\pi} \alpha d (A - a) \,. 
\end{equation}
The goal of this Appendix is to simplify this Lagrangian. Since $a$ has a Chern-Simons term at level 1, we can integrate out $a$ and obtain
\begin{equation}
    \begin{aligned}
    L &= \sum_{i=1}^3 \left\{L[\Psi_i, b_i] + \frac{2}{4\pi} \beta_i d \beta_i + \frac{1}{2\pi} \beta_i d b_i\right\} - \frac{2}{4\pi} \alpha d \alpha + \frac{1}{2\pi} \alpha d A - \frac{1}{4\pi} (\beta_1 + \beta_2 + \beta_3 + \alpha) d (\beta_1 + \beta_2 + \beta_3 + \alpha) \\
    &= \sum_{i=1}^3 L[\Psi_i, b_i] - \frac{3}{4\pi} \alpha d \alpha + \frac{1}{2\pi} \alpha d A + L_{\beta}
    \end{aligned}
\end{equation}
where $L_{\beta}$ collects all terms that depend on the $\beta_i$ gauge fields
\begin{equation}
    L_{\beta} = \frac{1}{4\pi} (\beta_1 d \beta_1 + \beta_2 d \beta_2 + \beta_3 d \beta_3) - \frac{1}{2\pi} (\beta_1 d \beta_2 + \beta_2 d \beta_3 + \beta_1 d \beta_3) + \sum_{i=1}^3 \frac{1}{2\pi} \beta_i d (b_i - \alpha) \,. 
\end{equation}
Integrating out $\beta_3$ gives
\begin{equation}
    \begin{aligned}
    L_{\beta} &= \frac{1}{4\pi} (\beta_1 d \beta_1 + \beta_2 d \beta_2) - \frac{1}{2\pi} \beta_1 d \beta_2 + \sum_{i=1}^2 \frac{1}{2\pi} \beta_i d (b_i - \alpha) - \frac{1}{4\pi} (\beta_1 + \beta_2 - b_3 + \alpha) d (\beta_1 + \beta_2 - b_3 + \alpha) - 2 \mathrm{CS}_g \\
    &= - \frac{2}{2\pi} \beta_1 d \beta_2 - \frac{1}{4\pi} (b_3 - \alpha) d (b_3 - \alpha) - 2 \mathrm{CS}_g + \sum_{i=1}^2 \frac{1}{2\pi} \beta_i d (b_i + b_3 - 2 \alpha) \,.  
    \end{aligned}
\end{equation}
Let us now plug this form of $L_{\beta}$ back into the full Lagrangian
\begin{equation}
    L = \sum_{i=1}^3 L[\Psi_i, b_i] - \frac{3}{4\pi} \alpha d \alpha + \frac{1}{2\pi} \alpha d A - \frac{2}{2\pi} \beta_1 d \beta_2 - \mathrm{CS}[b_3-\alpha,g] + \frac{1}{2\pi} \beta_1 d (b_1 + b_3 - 2 \alpha) + \frac{1}{2\pi} \beta_2 d (b_2 + b_3 - 2 \alpha) \,. 
\end{equation}
Integrating out $\beta_2$ allows us to make the substitution $b_3 = 2 \beta_1 + 2 \alpha - b_2$:
\begin{equation}
    \begin{aligned}
    L &= L[\Psi_1, b_1] + L[\Psi_2, b_2] + L[\Psi_3, 2 \beta_1 + 2 \alpha - b_2] - \frac{3}{4\pi} \alpha d \alpha + \frac{1}{2\pi} \alpha d A - \mathrm{CS}[2 \beta_1 + \alpha - b_2,g] + \frac{1}{2\pi} \beta_1 d (2 \beta_1 + b_1 - b_2) \\
    &= L[\Psi_1, b_1] + L[\Psi_2, b_2] + L[\Psi_3, 2 \beta_1 + 2 \alpha - b_2] - \frac{3}{4\pi} \alpha d \alpha + \frac{1}{2\pi} \alpha d A - \mathrm{CS}[b_2 - \alpha,g] + \frac{1}{2\pi} \beta_1 d (b_1 + b_2 - 2\alpha)
    \end{aligned}
\end{equation}
Shifting $\beta_1 \rightarrow \beta_1 - \alpha$ and integrating out $\alpha$ then gives
\begin{equation}
    \begin{aligned}
    L &= L[\Psi_1, b_1] + L[\Psi_2, b_2] + L[\Psi_3, 2 \beta_1 - b_2] - \mathrm{CS}[b_2,g] + \frac{1}{2\pi} \beta_1 d (b_1 + b_2) + \frac{1}{2\pi} \alpha d (A - b_1 - 2 \beta_1) \\
    &= L[\Psi_1, - A - 2 \beta_1] + L[\Psi_2, b_2] + L[\Psi_3, 2 \beta_1 - b_2] - \mathrm{CS}[b_2,g] + \frac{1}{2\pi} \beta_1 d (b_2 - A - 2 \beta_1) \,. 
    \end{aligned}
\end{equation}
As a consistency check, note that at zero doping, $\Psi_i$ are gapped and $b_2$ can be trivially integrated out. The remaining Lagrangian involving $\beta_1$ and $A$ is precisely the canonical Lagrangian for the Laughlin state at $\nu = 1/3$. 

\section{Competition between the flavor-symmetric and flavor-symmetry breaking paired states}\label{app:LG}

In this Appendix, we use a simple Landau-Ginzburg analysis to make some remarks about the competition between the flavor-symmetric paired state with $\Delta_{12} = \Delta_{13} = \Delta_{23}$ and the flavor-symmetry-breaking paired state with $\Delta_{12} \neq 0$ and $\Delta_{13} = \Delta_{23} = 0$. 

From the field theory in \eqref{eq:doped_simple}, it is clear that the effective interlayer pair fields $\Delta_{i \neq j}$ carry gauge charges under various dynamical gauge fields. Specifically, $\Delta_{12}$ is charged under $-A - 2 \beta_1 + b_2$, $\Delta_{13}$ is charged under $-A - b_2$, and $\Delta_{23}$ is charged under $2 \beta_1$. Up to quartic order, the allowed gauge-invariant combinations of these pair fields that can appear in the Lagrangian are therefore $|\Delta_{12}|^2, |\Delta_{13}|^2, |\Delta_{23}|^2$ and their products. The most general quartic Ginzburg-Landau theory then takes the form
\begin{equation}
    L = - r \left(|\Delta_{12}|^2 + |\Delta_{13}|^2 + |\Delta_{23}|^2\right) + u \left(|\Delta_{12}|^4 + |\Delta_{13}|^4 + |\Delta_{23}|^4\right) + v \left(|\Delta_{12}| \cdot |\Delta_{13}| + |\Delta_{12}| \cdot |\Delta_{23}| + |\Delta_{13}| \cdot |\Delta_{23}|\right)
\end{equation}
The first two terms describe three decoupled Landau-Ginzburg theories for each of the pairing fields $\Delta_{i\neq j}$, while the last term describes their competition. When $v > 0$/$v < 0$, different pair fields repel/attract each other. By minimizing the free energy subject to the positivity constraints, we see that 
\begin{enumerate}
    \item For $v < 2 u$, the repulsion between different pair fields is weak. In this regime, the pair fields prefer to preserve the $\mathbb{Z}_3$ flavor symmetry. 
    \item For $v > 2 u$, the repulsion between pair fields is so strong that once one of them onsets, the other two prefer to vanish. This fully polarized limit is the maximally flavor-symmetry breaking state. 
\end{enumerate}
From this analysis, we see that up to quartic order, there is no parameter regime in which the pair fields partially polarize. Partial polarization can arise when higher-order interactions are included. However, so long as the partially polarized pair field configuration fully gaps the Fermi surfaces, the BdG Chern number is the same as the flavor-symmetric case. Therefore, universal properties of the resulting electronic superconductor will also agree with the flavor-symmetric case. 

\section{Orthogonal metal from an $SU(3)$ symmetric ansatz}
\label{app: z3om} 

Assuming an approximate $SU(3)$ symmetry at low doping, it is natural to consider a parton ansatz that preserves the $SU(3)$ symmetry. Towards that end, we fractionalize the CF valleys as $f_i = \chi_i \Phi$ where $\chi_i$ are fermionic partons and $\Phi$ is a hard-core boson
\begin{equation}
    \sum_{i=1}^3 L[f_i, a] = \sum_{i=1}^3 L[\chi_i, b] + L[\Phi, a - b] \,.
\end{equation}
If we choose the mean-field flux of $b$ to vanish, then $\Phi$ is at filling $\nu_{\Phi} = -3/2$ while $\chi_i$ sees zero magnetic field on average. The simplest mean-field that preserves $SU(3)$ symmetry is to have $\chi_i$ form a $SU(3)$ symmetric Fermi liquid and $\Phi$ form a bosonic integer quantum Hall state at $\nu_{\Phi} = -2$ stacked with a bosonic Laughlin state at $\nu_{\Phi} = 1/2$. Combining with the rest of the Lagrangian gives 
\begin{equation}
    \begin{aligned}
    L &= \frac{1}{4\pi} a da + 2 \mathrm{CS}_g - \frac{2}{4\pi} \alpha d \alpha + \frac{1}{2\pi} \alpha d (A-a) + \sum_{i=1}^3 L[\chi_i, b] - \frac{2}{4\pi} (a-b) d (a-b) - \frac{2}{4\pi} \beta d \beta + \frac{1}{2\pi} \beta d (a-b) \\
    &= \sum_{i=1}^3 L[\chi_i, b] - \frac{2}{4\pi} \alpha d \alpha + \frac{1}{2\pi} \alpha d A - \frac{2}{4\pi} b db + 4 \mathrm{CS}_g - \frac{2}{4\pi} \beta d \beta - \frac{1}{2\pi} \beta d b - \frac{1}{4\pi} a da - 2 \mathrm{CS}_g + \frac{1}{2\pi} a d (\beta - \alpha + 2b)  \,.
    \end{aligned}
\end{equation}
Integrating out $a$ gives
\begin{equation}
    L = \sum_{i=1}^3 L[\chi_i, b] - \frac{2}{4\pi} \alpha d \alpha + \frac{1}{2\pi} \alpha d A - \frac{2}{4\pi} b db + 4 \mathrm{CS}_g - \frac{2}{4\pi} \beta d \beta - \frac{1}{2\pi} \beta d b + \frac{1}{4\pi} (\beta - \alpha + 2b) d (\beta - \alpha + 2b) \,. 
\end{equation}
Further integrating out $\beta$ gives
\begin{equation}
    \begin{aligned}
    L &= \sum_{i=1}^3 L[\chi_i, b] - \frac{2}{4\pi} \alpha d \alpha + \frac{1}{2\pi} \alpha d A - \frac{2}{4\pi} b db + 4 \mathrm{CS}_g + \frac{1}{4\pi} (2b - \alpha) d (2b-\alpha) + \frac{1}{4\pi} (b- \alpha) d (b-\alpha) + 2 \mathrm{CS}_g \\
    &= \sum_{i=1}^3 L[\chi_i, b] + \frac{1}{2\pi} \alpha d (A - 3 b) + 3 \mathrm{CS}[b,g] \,.  
    \end{aligned}
\end{equation}
The integral over $\alpha$ sets $A = 3b$. In other words, $b$ is Higgsed down to a $\mathbb{Z}_3$ gauge field. The final Lagrangian therefore describes three identical Fermi surfaces coupled to a common $\mathbb{Z}_3$ dynamical gauge field. This phase resembles the $\mathbb{Z}_3$ orthogonal metal~\cite{nandkishore2012orthogonal} but has a crucial difference: the electron spectral function remains gapless, although it has an incoherent structure without a quasiparticle-like peak at the Fermi surface. Various other properties of this state follow directly from the Lagrangian description, as discussed in the main text.

\end{document}